This is the pre-peer reviewed version of the following article: Verma, A. and Sharma, A. , Enhanced Self-Organized Dewetting of Ultrathin Polymer Films Under Water-Organic Solutions: Fabrication of Sub-micrometer Spherical Lens Arrays. *Advanced Materials*, n/a. doi: 10.1002/adma.201002768, which has been published in final form at: http://onlinelibrary.wiley.com/doi/10.1002/adma.201002768/abstract

**Enhanced Self-organized Dewetting of Ultrathin Polymer Films under Water-organic Solutions: Fabrication of Sub-micron Spherical Lens Arrays**

By *Ankur Verma* and *Ashutosh Sharma**

Department of Chemical Engineering & DST unit on Nanosciences
Indian Institute of Technology Kanpur, Kanpur 208016, India
E-mail: ashutos@iitk.ac.in

Dewetting of ultrathin (< 100 nm) polymer films and their self-organization on physico-chemically patterned substrates has been extensively studied with potential applications in soft-patterning.[1] However, despite its scientific and technological promise, there are two major limitations of the self-assembled patterns: (1) The length scale of the spinodal instability and thus the size of the resulting features are severely limited by the surface tension, $\gamma$ effects on small scales. The wavelength of the long-wave instability is given by:[2] $\lambda = [-8\pi^2\gamma/(\partial\phi/\partial h)]^{1/2}$, where $h$ is the film thickness and $\phi$ is the destabilizing potential ($\sim h^{-3}$ for van der Waals attraction). For example, dewetting of films in the range of 20 nm to 100 nm produce feature sizes in the range of 5 μm to 100 μm[1,2]; (2) The micro-structures produced by the long-wave instability have very small aspect-ratios and small contact angles in air.[1,2]

We propose here a very simple method to fabricate sub-micron (~ 100 nm) ordered and tunable polymeric structures by self-organized room temperature dewetting of ultrathin polymer films by minimizing the surface tension limitation. The method is illustrated by fabrication of sub-micron size polymeric lens arrays where the lens curvature can be easily controlled. The technique proposed here induces a room temperature dewetting of polystyrene thin films under an optimal mix of water, acetone and methyl-ethyl ketone (MEK). Selective permeation of the organic solvents in the polymer reduces its glass transition temperature below the room temperature and also greatly decreases the interfacial tension, without a concurrent solubilization of PS owing to water being the majority phase in the outside mixture. In addition, we also show that dewetting under MEK mixture accelerates the kinetics of dewetting and more importantly, greatly increases the contact angle of sub-micron droplets compared to dewetting in air. Finally, we illustrate the technique by fabricating an ordered array of sub-micron lenses (~ 200 nm diameter) of tunable curvature.

As regards the context of micro/nano lenses, diffraction limit (Ernst Abbe, 1873) prevents resolving two objects closer than half the wavelength of light using lens based far-field optics. For a finer optical resolution, one has to employ lensing effect by surface-plasmon excitation[3] or fluorescence microscopy involving molecular excitations.[4] Until very recently, the only way to see sub-diffraction limited features was by using near-field optical microscope which captures these evanescent waves by nanosized mechanical tip. In recent years there are various efforts to recover and project the evanescent waves into the far field. Metamaterials based superlens with properties like negative refractive index offer one possible solution to overcome the diffraction limit.[5] Use of spherical micro- and nano-lenses offers new possibilities of lens based near-field detection and high resolution optical imaging at low intensities. These micro-lenses are shown to provide lens based near-field focusing in high resolution optical microscopy.[6] In biological systems there are several examples of array of miniaturized lenses such as insect's compound eyes.[7] Micro-lenses are used in a variety of optical devices like charge-coupled device (CCD), digital projectors and photovoltaics. Fabrication of such micro and nano-lenses is carried out mainly by top down approach using sophisticated tools like lithography, contact printing, inkjet printing, focused ion beam, and UV laser etc., but there are recent efforts at the use of self-assembly for fabricating a single lens.[8]

We investigate here another approach for the fabrication of a single micro/nano lens as well as lens arrays based on controlled dewetting of ultrathin polymer films which has been studied extensively over the past two decades.[1–2,9–10] When a polymer thin-film (thickness <100 nm) is heated above its glass transition temperature ($T_g$) or $T_g$ is decreased by solvent vapor exposure, the inter-surface interactions such as the van der Waals engender the formation of holes which grow and coalesce to eventually form isolated spherical droplets. The size and the mean distance between these droplets are usually in the range of 10s of microns and are functions of the film thickness. Further the shape of the droplet depends on the interfacial tension or the equilibrium contact angle. These smooth droplets when quenched to their glassy state can act like micro-lenses. However, dewetting in air produces relatively large droplets of small contact angle which are randomly distributed on the substrate.[2] More ordered assemblies can be formed on physico-chemically structured substrates.[1,10] As shown below, the first two limitations on the shape and size are overcome by dewetting under MEK mixture and the ordered arrays are facilitated by controlled dewetting on patterned substrates that precisely control the spatial locations of dewetting and the placement of dewetted structures. Dewetting in air on physically and chemically patterned substrates has been studied both experimentally and theoretically.[1,9–10]

Another challenge in micro-lens fabrication is to tailor the equilibrium contact angle of these droplets to obtain different focal lengths. We show that dewetted droplets under the MEK mix slowly increase their contact angle thus allowing for the possibility of controlling the lens curvature. The solvent molecules go inside the polymer matrix allowing the polymer chains to glide over one another and thus bringing its $T_g$ below room temperature. However the presence of water, a highly polar molecule, inhibits the solubility of polymer in the dewetting mixture. Thus, dewetting in the liquid media adds the benefit of cleaner environment, better control and better possibility of defect control compared to thermal annealing.

On a flat homogeneous surface, dewetting of a thin polymer film is engendered by the excess intermolecular forces.[1–2,11] **Figure 1** shows the sequence of optical images taken at different intervals to show various stages of dewetting for a 25 nm thick polystyrene (PS) film in the MEK mixture. Dewetting starts with the formation of randomly distributed isolated dry spots or holes on the film, which in time grow and coalesce to form a network of polymers that breaks in to isolated droplets. For this film, the onset of instability was observed after 10 seconds of placing it inside the immersing liquid which is significantly shorter than the time (several minutes) taken for dewetting in air. Nearly uniform sized holes can be seen on the film after 30 seconds (Figure 1a). The number of holes per unit area, $N$ displayed approximately the scaling, $N \sim \lambda^{-2} \sim h^{-4}$ (results not shown). This resembles the theoretical scaling $\lambda \sim [(\partial\phi/\partial h)]^{-1/2}$ obtained for the van der Waals potential, $\phi \sim h^{-3}$. Regardless of the precise mechanism of hole-formation, this scaling is same as seen in most of the previous studies.[1,2] These holes grow in size and subsequently coalesce in 6 minutes to form a network of polymer (Figure 1b, 1c). Within 15 minutes this network of polymer breaks into isolated droplets to complete the dewetting (Figure 1d). After this stage there is no further change in the overall morphology, but the contact-lines of individual droplets continue to slowly recede for about one hour. This decrease in the droplet base-diameter is accompanied by an increase in the equilibrium contact angle and the aspect ratio of the droplet. The slow kinetics of these droplets is because of their high viscosity, which allows for a facile control of lens curvature by freezing a desired intermediate structure by removal from the MEK medium. Droplets of much smaller molecular weight PS retracted to the equilibrium shape much faster within a couple of minutes (not shown). The number density of droplets in the fully dewetted sample of 25 nm thick PS film is found to be $3.9\times10^{4}$ mm$^{-2}$ that gives the mean separation of 5.1±0.2 μm between droplets. The corresponding value for dewetting in air is 24.2±1.9 μm.

Mean droplet diameter obtained in MEK mixture is significantly smaller than in air. **Figure 2** compares the droplet diameters, $d$ for both the cases as a function of film thickness, which clearly shows nearly one order reduction in the droplet size. Moreover, the size dependence of the diameter on film thickness is also weaker for dewetting under MEK mix, viz. $d \sim h^{-1.17\pm0.09}$ as opposed to $h^{-1.49\pm0.05}$ for dewetting in air. The latter scaling in air is well understood in the works of Reiter[2] and Sharma and Reiter.[2] The size-reduction under MEK mix is engendered by a decrease in the interfacial tension owing to the presence of the solvent and also because of anomalously strong destabilizing forces observed for thin liquid polymer films under water.[11] A weaker dependence on the film thickness, $d \sim h^{-1.17}$ is consistent with a reduced exponent of the dominant electrostatic destabilizing force under water ($\phi\sim h^{-2.5}$)[11] as compared to the van der Waals force in air ($\phi\sim h^{-3}$).

The dewetted droplets were examined by placing the substrate vertically in the field emission scanning electron microscope (FESEM) to see the side view of these droplets. **Figure 3** (a–c) shows evolution of contact angles of PS droplets with time. The contact angle slowly increases from 83° after 15 minutes of dewetting to about 140° after 1 hour. Figure 3 (d–l) illustrates a variety of lenses obtained by dewetting of 14-60 nm thick films producing droplets with the diameter and contact-angle in the range of 370 nm to 2.9 μm and 30°–143°, respectively. It is evident from the figure that a significantly wide range of micro-lenses both in size and focal length can be fabricated using this method. All these structures are stable in air at room temperature. Contact angles of less than 40° were only observed in smaller droplets with a diameter less than 800 nm. However, larger sized droplets with smaller contact angle can be made by heating them above their $T_g$ in air. This shape change is completely reversible and once dried, droplet with all size and shape are stable at room temperature. This reversible shape changing capability of these microlenses adds to their flexibility to be used in optical devices.

In order to obtain a uniform array of lenses, the dewetting of PS thin films was carried out on a pre-patterned substrate with a 2-D array of cylindrical pillars. **Figure 4** illustrates an example of the fully dewetted structure of a 25 nm thick PS film on a physically patterned substrate. In Figure 3a the brighter spots are the electron-beam lithography (EBL) fabricated cylindrical pillars, while the intervening spots are the polymer droplets. The effect of pattern can be observed clearly as the alignment is seen only in the regions where the pre-fabricated pillars are present. 2-D confinement also further reduces the average size of droplets as well as the mean wavelength. In the transverse FESEM image the array of polymer droplets can be seen (Figure 4b).

To see the lensing effect of fabricated PS microlenses, optical characterization was performed by putting glass coverslip containing microlenses over the object to be seen in optical microscope. **Figure 5a** displays the scheme of the arrangement, where micro-lenses are placed between the object and the objective lens of the microscope. The micro-lenses used here are 3−7 μm in size with contact angle close to 70°. It is evident from the images that the presence of microlenses adds to the resolution of the microscope. The pattern below the microlens resolves better than the regions without microlens. The effectiveness of PS microlenses in improving the resolution is evident with 500 nm wide parallel strips on silicon wafer (Figure 5b) and 800 nm wide parallel tracks on a compact disk (CD) (Figure 5c).

Dewetting of polymer ultrathin films under water and MEK mix thus proves to be a powerful but flexible, simple and inexpensive technique for the room temperature fabrication of sub-micron polymeric structures and spherical lenses by physical self-assembly. Further, self-organized dewetting of polymer thin films on a physically patterned surface is shown to produce a regular 2-D array of microlenses over a large area of the order of $cm^2$. The presence of water as the bounding fluid increses the contact angle of the droplet, whereas the heating in air can decrease the contact angle. This change is completely reversible and can be used to make as stimuli-responsive microlenses capable of changing their focal lengths when subjected to appropriate conditions. Lensing effect of these microlenses is also demonstrated and they prove to be a promising in resolving smaller objects in a optical microscope.

*Experimental*

Polystyrene (PS) of average molecular weight ($M_w$) 280,000 g mol$^{-1}$ was purchased from Sigma Aldrich and used as it is. Polydispersity index (PDI = $M_w/M_n$) of PS was less than 1.1. Thin-films of thickness ranging from 10 nm to 60 nm were spin coated at 3000 rpm on thoroughly cleaned silicon wafers with an oxide layer (~ 2−3 nm measured by ellipsometry) by using 0.1–1 w/v % polymer solution in HPLC grade toluene. After spin coating, the films were dried in air for 2 hours and annealed in a low vacuum oven at 60°C for 12 hours to minimize residual stresses developed during spin coating. 2-D array of cylindrical pillars of diameter 250 nm, height 100 nm and pitch 1 μm were made on silicon by EBL. EBL patterned silicon wafers were directly used for spin coating of PS films. In order to test the optical performance of microlenses, PS films were coated on cleaned glass coverslips and dewetted structures on these transparent substrates were then used for imaging after annealing at 110°C for 30 minutes. Annealing removed the residual solvents and also ensured contact angle of less than 90°.

Annealed thin-films of PS were put in dewetting chamber containing liquid mixture of water, acetone and methyl-ethyl ketone (MEK) in the ratio 15:3:7. Composition of this mixture is an important parameter to tune the interfacial tension. MEK being the good solvent of PS it effectively reduces its glass transition temperature and allows the dewetting to occur at room temperature. This mixture has no significant solubility for PS. The samples were taken out at different intervals to examine the extent of dewetting and fully dewetted samples were taken for the isolation of dewetted structures.

Thickness measurements of spin coated thin films were carried out by nulling ellipsometer (Nanofilm, EP$^3$-SE) using 532 nm green laser at an incident angle of 55°. Refractive index of PS is taken as the value for bulk, i.e. 1.58. Imaging of the fabricated structures was done using optical microscope (Zeiss Axio observer Z1) in bright field and FESEM (Zeiss Supra 40VP) using secondary electrons. To measure the contact angle of polymer droplets substrates were examined vertically in FESEM to see the transverse view of droplets. Optical characterization of fabricated microlenses was done using optical microscopy.

*Acknowledgements*

This work is supported by Department of Science and Technology unit on Nanosciences and by an IRHPA grant. Authors acknowledge the initial efforts of Rabibrata Mukherjee and Partho Pattader in setting up some of the experiments.

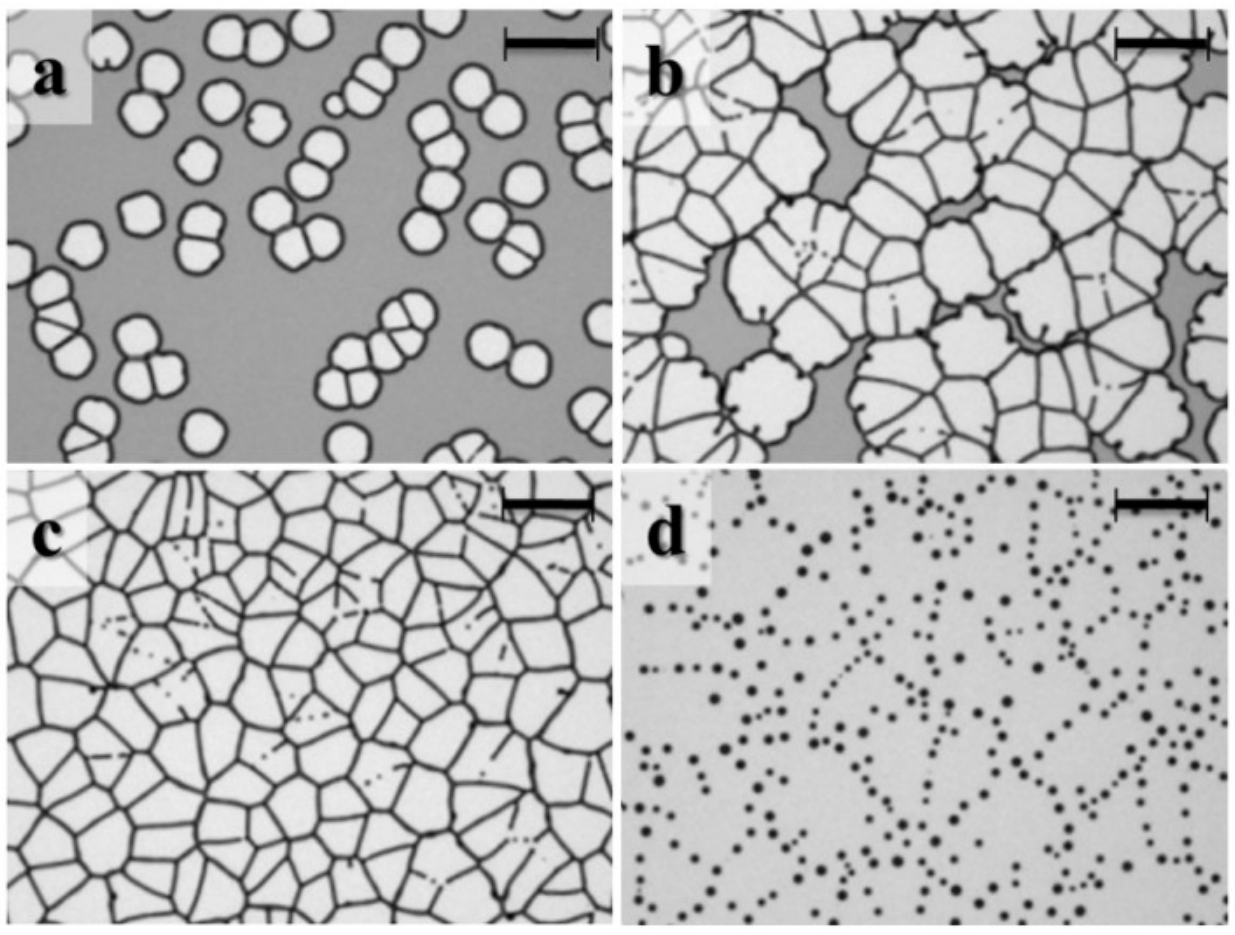

**Figure 1.** Time evolution of dewetting of a 25 nm PS film on a flat silicon substrate under MEK solution. Morphology of PS thin-film after (a) 30 seconds as randomly distributed isolated holes, (b) 2 minutes as partially coalesced holes, (c) 6 minutes as completely coalesced holes forming a network and (d) 15 minutes as isolated polymer droplets. The number density of droplets in the fully dewetted sample is $3.9 \times 10^4$ /mm$^2$ and the mean separation between droplets to be 5.1 μm. The corresponding value for in-air dewetting is 24 μm. (Scale bar: 10 μm)

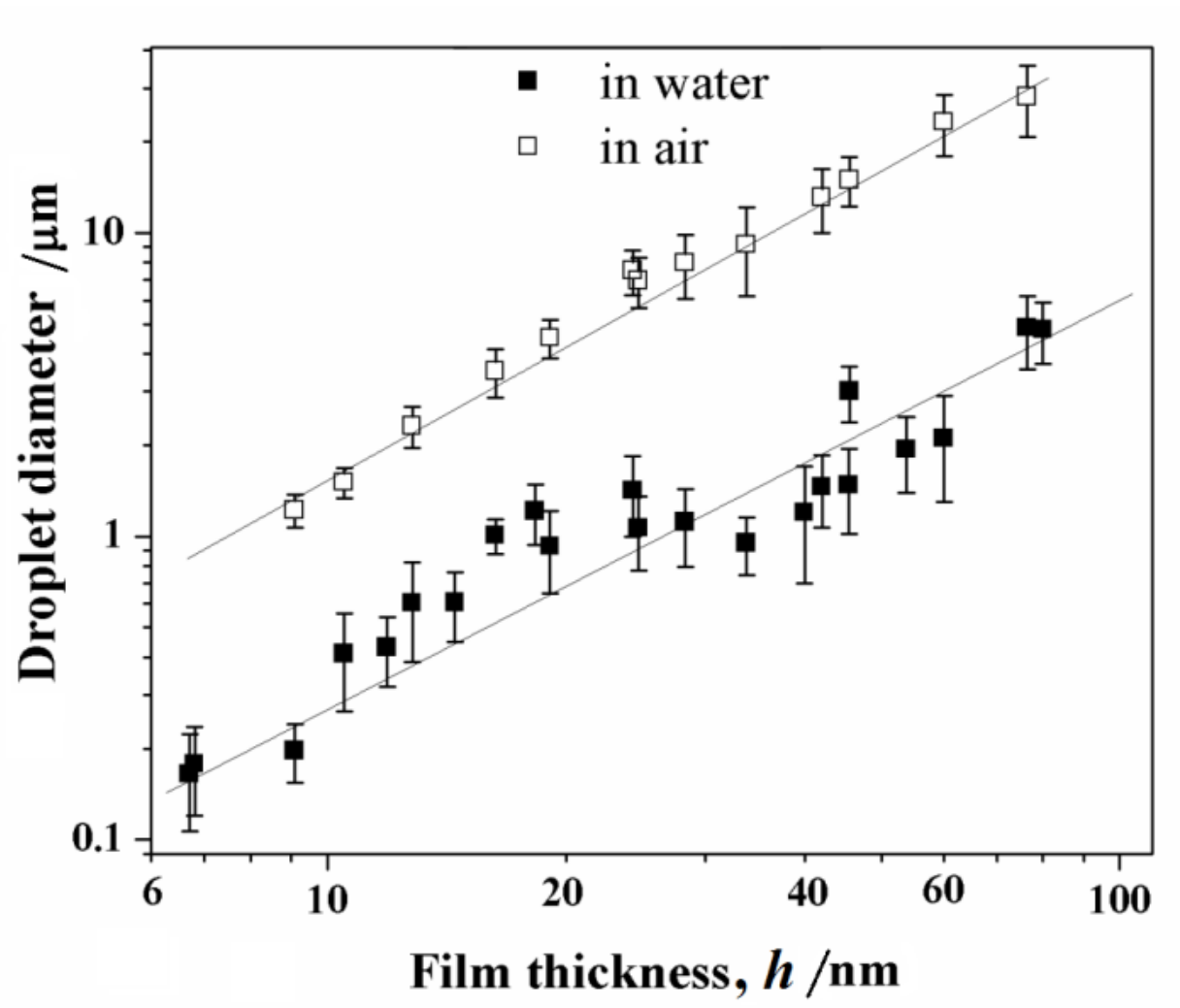

**Figure 2.** Comparison of the average droplet diameters for dewetting in air and in water. Droplet size is nearly one order of magnitude smaller in the case of dewetting under the water-MEK mixture. Slope of the best fit lines on the log-log plot gives the film thickness ($h$) dependence of droplet size as $h^{-1.17\pm0.09}$ for dewetting under MEK as opposed to $h^{-1.49\pm0.05}$ for dewetting in air.

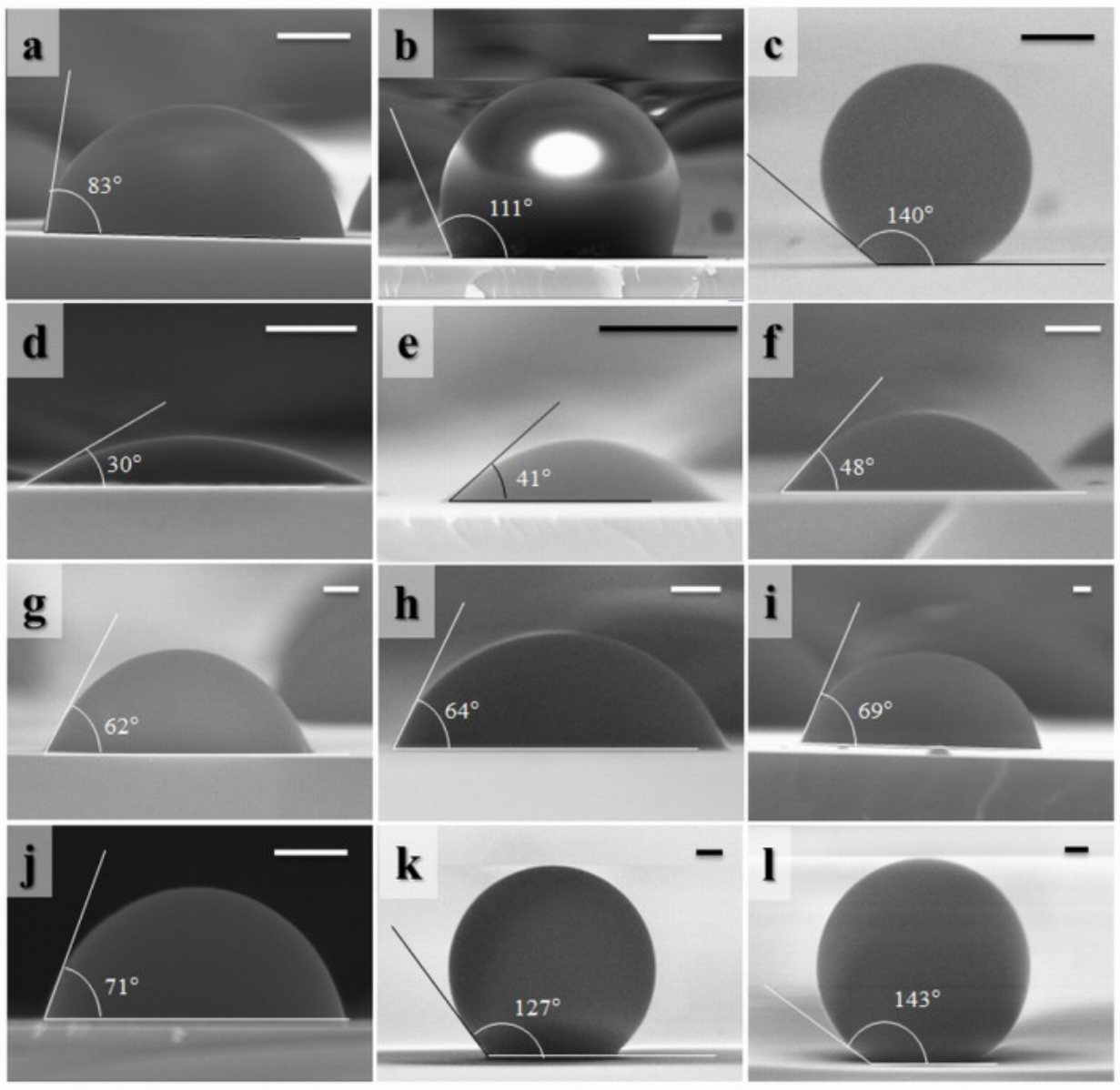

**Figure 3.** Transverse FESEM images of PS droplets. Images (a) – (c) show time evolution of a droplet obtained by dewetting of a 25 nm thick PS film on a flat silicon substrate. Contact angle after: (a) 15 min is 83°, (b) 20 min is 111°, and (c) 1 hour is 140°. Images (d) – (l) show a range of droplet shapes and sizes obtained by dewetting of 14–60 nm thick films. Their base-radii are in the range from 370 nm to 2.9 μm and the contact angle varies from 30° to 143°. (Scale bar: 200 nm)

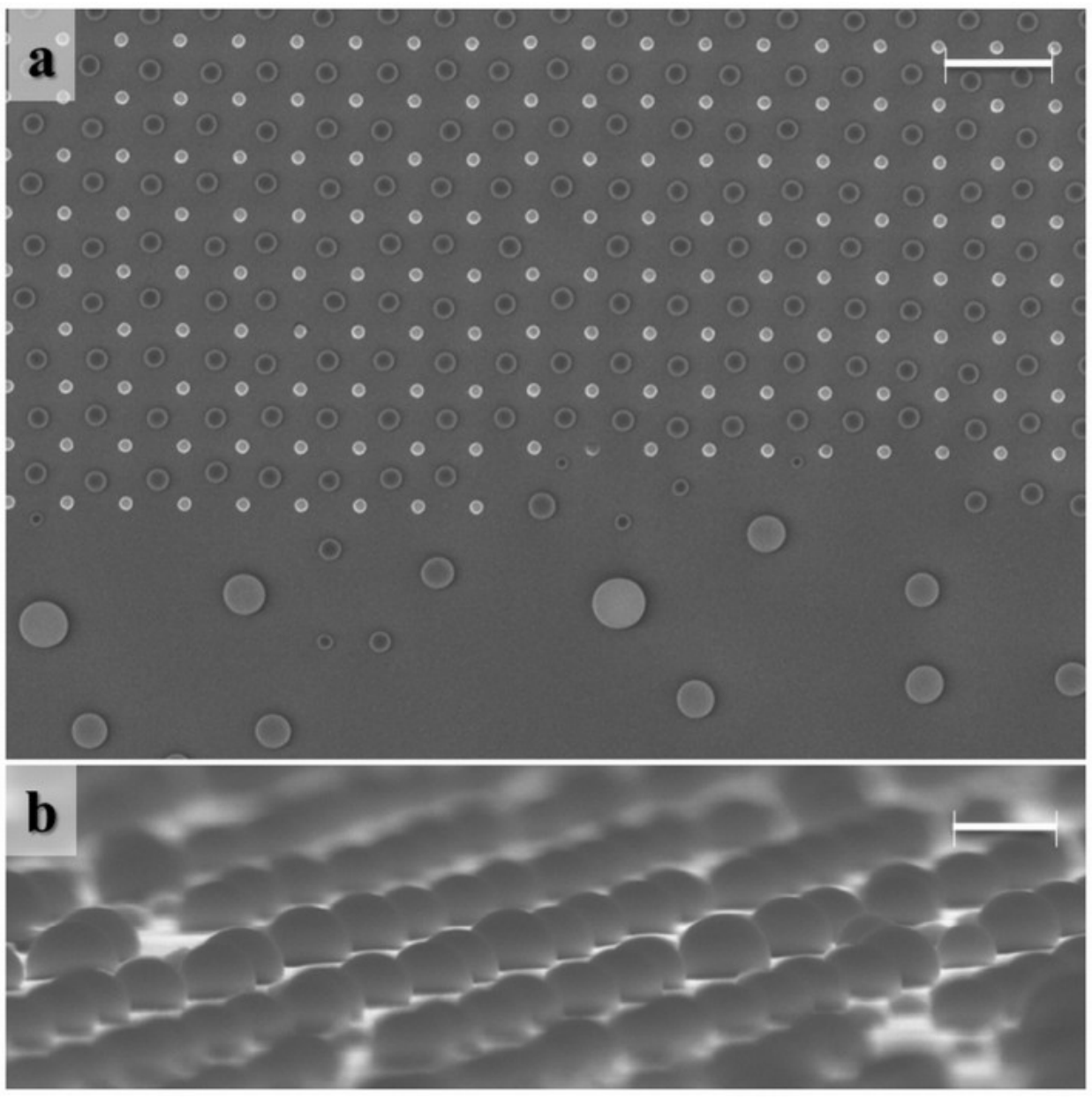

**Figure 4.** Dewetting on a physically patterned substrate (a) FESEM image of the dewetted structure (Scale bar: 2 μm), (b) Transverse view showing arrays of dewetted droplets. (Scale bar: 500 nm)

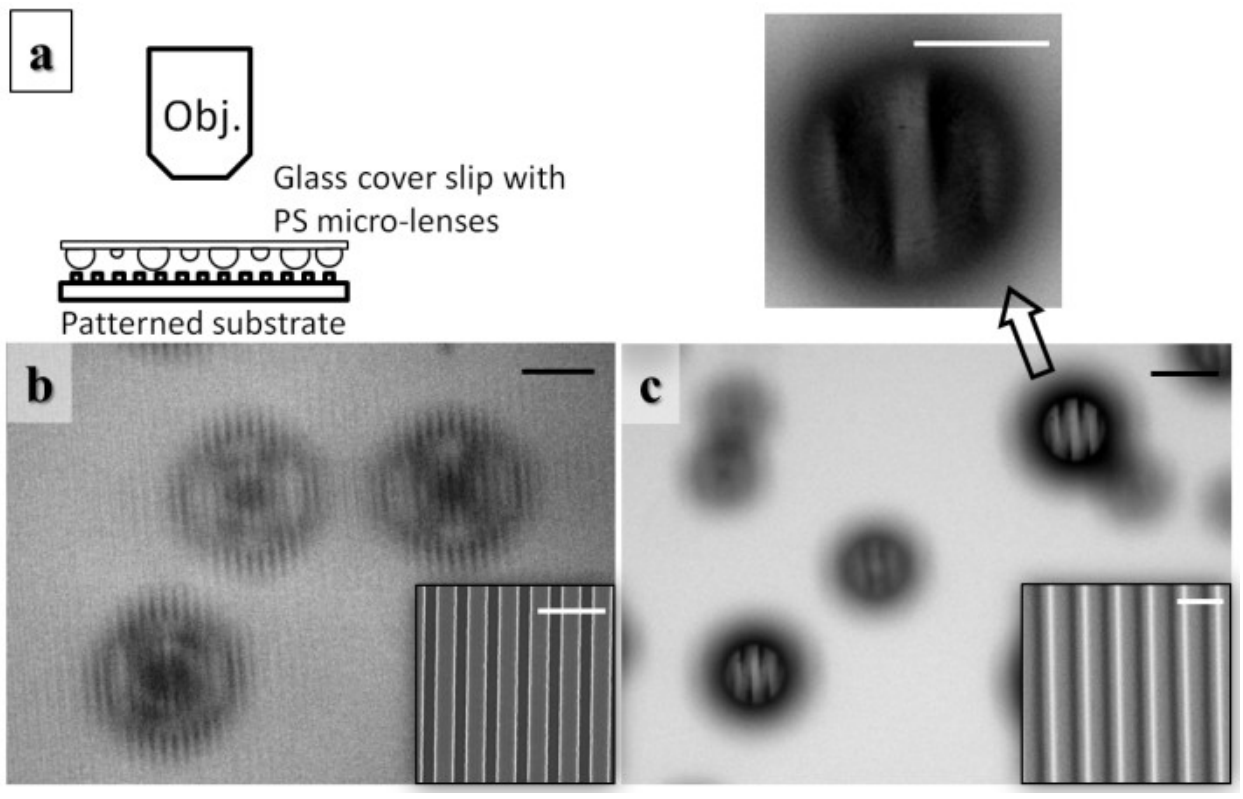

**Figure 5.** Resolving striped patterns in optical microscope by PS micro-lenses. (a) Schematic diagram of visualization using optical microscope. (b) 500 nm wide strips on silicon wafer resolved by 50X objective (NA/0.5), inset shows the FESEM image of the object. (c) CD strips with channel width of 800 nm resolved by 20X (NA/0.4) objective, enlarged view is of 50X objective, inset shows the FESEM image of the object. Scale bar: 5 μm (black), 2 μm (white)